\documentclass[aps,preprint,showpacs,superscriptaddress,groupedaddress]{revtex4}  
\usepackage{graphicx}  
\usepackage{dcolumn}   
\usepackage{bm}        
\usepackage{amssymb}   
\begin{document}


\title{Hadronic interaction models and the angular distribution of cosmic ray muons}
\author{Dimitra Atri}
\affiliation{Particle Astrophysics Group, Blue Marble Space Institute of Science\\Seattle, WA 98145-1561, USA}
\affiliation{Tata Institute of Fundamental Research\\Colaba, Mumbai 400 005, India\\dimitra@bmsis.org}

\begin{abstract}

Muons serve as the best probes of the physics of hadronic interactions in the upper atmosphere because of their simple physics. All the properties of detected muons, such as their energy and angle of incidence, are governed by the properties of their parent hadrons. The angular distribution of the detected muons is a result of a superposition of multiple effects, and is governed by a number of parameters, such as the cutoff rigidity (magnetic field), atmospheric attenuation, interaction cross sections, production height, multiplicity and pseudorapidity. Since particle interactions are handled by the low and high-energy hadronic interaction models in air shower simulations, subtle differences in their physics should manifest in the differences in the angular distribution. Here, we compare air shower simulations with experimental data and investigate how the choice of hadronic interaction models in air shower simulations affects the angular distribution of cosmic ray muons. 
 
\end{abstract}

\maketitle

\section{Introduction}
Extensive air showers are produced when high energy particles of astrophysical origin interact with the Earth's atmosphere. Particles produced in the air shower are primarily a result of the particle-air hadronic interactions and subsequently electromagnetic interactions as the particles traverse in the atmosphere. Muons are byproducts of these hadronic interactions occurring in the upper atmosphere. Muons only undergo electromagnetic interactions and reach the detector without much loss in valuable information about the interactions due to relatively long half-life and small hadronic cross section. Muons are thus, the best messengers of hadronic interactions in the upper atmosphere. 

The total number of muons observed in air shower experiments, $n_{\mu}$ is an important parameter, and is widely used to differentiate between the interaction models \cite{tanguy2}. This differentiation is very important because all observations in cosmic ray physics, such as the cosmic ray composition, spectrum etc., are model dependent. Several studies have compared experimental data with the results of interaction models to identify the differences in observed and simulated $n_{\mu}$ \cite{kascade1}\cite{kascade2}. Such differences in the observed and predicted total number of muons are also easy to see in air shower experiments. Understanding the angular distribution of cosmic ray muons on the other hand is more tricky because it is a result of a superposition of multiple effects. 

The angular distribution primarily depends on the bending of charged particles in the geomagnetic field. Regions closer to the equator experience a significant East-West asymmetry in cosmic ray intensity due to a significant asymmetry in the cutoff rigidity. Positive particles from western directions are subjected to a lower rigidity cutoff than from the east. Since cosmic ray primaries are mostly positive particles, more particles are incident from the western directions than from the east. As a result of this cutoff rigidity, the energy threshold of positive particles incident from East is also higher than for the particles from West. This leads to different mean energies of primaries incident from different directions. The geomagnetic field, thus, serves as a natural magnetic spectrometer. Since the daughter products of hadronic interactions primarily depend on the energy of the primary particle, this East-West asymmetry is also present in the cosmic ray muon data observed at the ground level. Directional dependency of geomagnetic cutoff rigidity thus allows us to study primaries of different energy. For example, muons detected in Eastern direction have systematically higher energy and are produced by primaries of higher energy, in comparison to muons from West. 

Another factor affecting the distribution, is the cosmic ray muon charge ratio. The muon charge ratio $ R_{\mu} = (\mu^{+}/\mu^{-}) $, is important here because opposite charges bend in opposite ways in the magnetic field (Table 4). It must be noted here that air shower particles also bend significantly (depending on the energy) as they traverse in the atmosphere due to the geomagnetic field \cite{lipari}. Changes in this ratio can change the bending of muons and such changes would also be reflected in the muon distribution. Atmospheric attenuation also plays an important role and gives less number of muons at larger zenith angles. 

All hadronic interaction models have energy dependent cross sections, multiplicities, pseudorapidities and muon charge ratios \cite{tanguy1}\cite{sibyll}\cite{qgs}. These small differences in models, which are energy dependent, should be evident in computer simulations of sufficiently high statistics. However, it must be noted that the particles detected on the ground are a result of multiple interactions in the atmosphere. Shower to shower fluctuations are physically impossible to avoid due to stochastic nature of particle interactions and shower development. For example, the angle of incidence of muons is also governed by their production height, along with the angle of incidence of the primary particle. The production height in turn, depends on hadronic cross section, which is again a distribution, rather than a fixed point. It has been shown using a tracking muon detector, that the production height of muons can be determined, which can give insights into the hadronic interaction processes in the upper atmosphere \cite{kmuon}. Similarly, the angular distribution can be used to identify differences in different models, not necessarily a single parameter such as muon multiplicity or production height. Angular distribution is a result of a number of parameters such as the magnetic field, cutoff rigidity, atmospheric attenuation, cross sections (production height), multiplicity and pseudorapidity \cite{kmuon2}. Since particle interactions are handled by the low and high-energy hadronic interaction models in air shower simulations, subtle differences in their output should manifest in the differences in the angular distribution.

It must be mentioned here that the primary composition is a very important factor determining the muon angular distribution. However, in order to study the differences between hadronic interaction models, this component has been purposefully discarded and only proton primaries have been used in this work.

\section{Method}
\subsection{Modeling the geomagnetic field}
Since both primary and secondary particles bend in presence of the magnetic field, an accurate model of the geomagnetic field is crucial. The International Geomagnetic Reference Field (IGRF) 11th generation model \cite{finlay2010international}, which is a gold standard in modeling the geomagnetic field, has been used for this work. The code is a series of mathematical models of the Earth's magnetic field and its secular variations. It is a widely used code in different areas ranging from geosciences, atmospheric sciences to astroparticle physics. Once the geomagnetic model is well defined with IGRF, cosmic ray trajectories need to be calculated for the purpose of computing the cutoff rigidities of incident primaries. Particles with higher momentum undergo less bending than lower momentum particles for a given magnetic field. Cosmic ray trajectories were computed using the Multi-platform cosmic-ray trajectory program \cite{smart1}. The program is validated against real data \cite{smart2} and further details can be found therein. This code provides with the allowed and forbidden trajectories for positively charged particles with IGRF 11 magnetic field model. Trajectories are calculated using the backtracking technique where an antiproton of a particular momentum is launched towards space and its trajectory is traced. If it bends enough such that it falls back on the Earth, it's trajectory is marked as ``forbidden" because a proton of the same momentum incident from space will never be able to reach the Earth's atmosphere. Similarly, for antiprotons which are able to escape into space are marked as ``allowed". The magnetic field filters out lower energy primaries and only sufficiently energetic particles are able to reach the ``top" of the Earth's atmosphere. We then sort this data and compute the angular and azimuthal dependence of cutoff rigidities at Ooty, India, located at $11.4^{o}$N latitude, $76.7^{o}$E and 2200 m altitude above mean sea level. 

Due to the proximity of the experiment to the equator, a considerable azimuthal asymmetry in the cutoff rigidity is observed. Figure 1 shows the rigidity cutoff profiles in different directions at different zenith angles. As expected, primaries incident at higher zenith angles are subjected to a higher cutoff. Also, a significant East-West asymmetry is clearly visible.  

\begin{figure}[h] \begin{center}
\includegraphics[width=6.5in,height=5.0in]{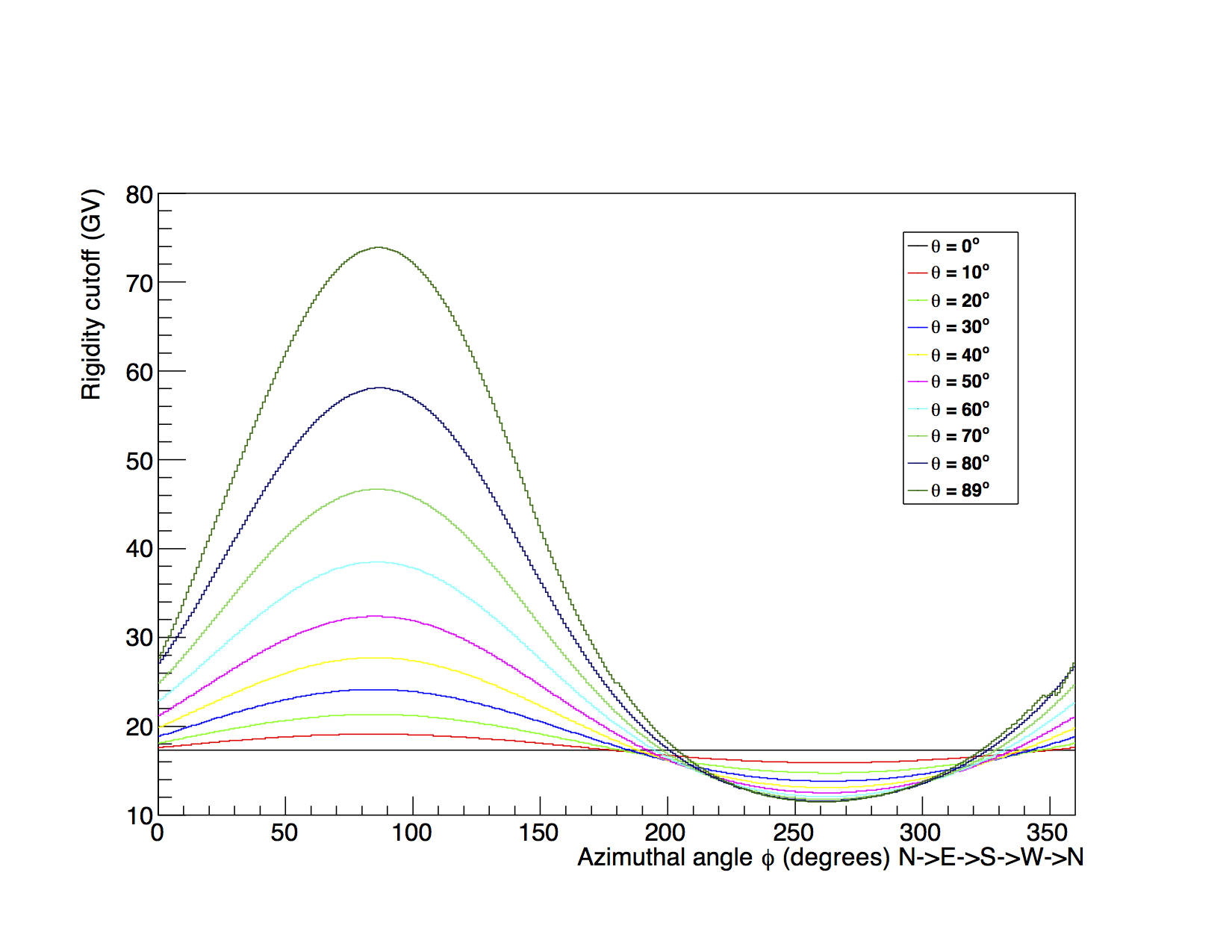}
\caption{\label{fig1} Rigidity cutoff at different zenith and azimuth angles. $0^{o}$ azimuth corresponds to North, $90^{o}$ is East and so on.}
\end{center} \end{figure} \medskip

\subsection{The muon detector arrangement}
\begin{figure}[h] \begin{center}
\includegraphics[width=6.75in,height=1.75in]{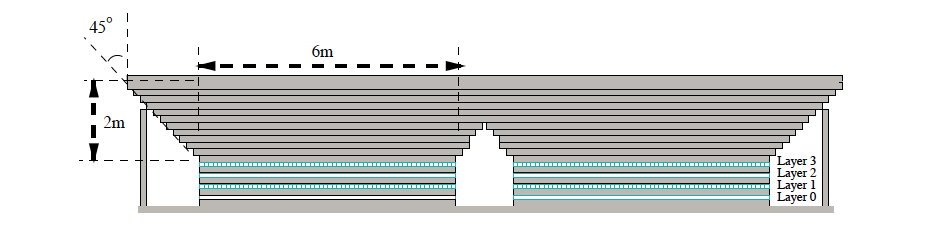}
\caption{\label{fig2} Schematic of a muon detector module showing four layers of PRCs embedded within the concrete structure. Two adjacent modules are shown, separated by a horizontal distance of 130 cm at the base.}
\end{center} \end{figure} \medskip

\begin{figure}[h] \begin{center}
\includegraphics[width=4.0in,height=3in]{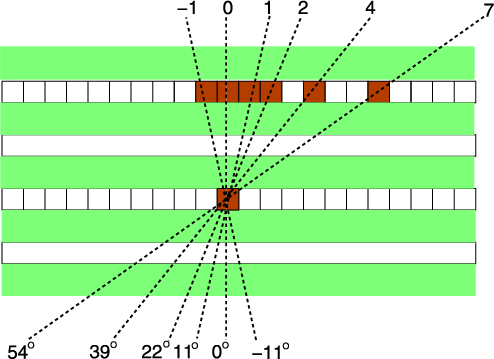}
\caption{\label{fig3} Angular reconstruction of muon tracks in a module}
\end{center} \end{figure} \medskip
The muon detector arrangement comprise of 16 modules of 35 $m^{2}$ of sensitive area in each, and a total area of 560 $m^{2}$. The arrangement is shown in detail in Figure 2, 3 and 4. The basic element of the detector is a gas proportional counter (PRC), each of which is 6 m long and has a cross sectional area of 10 cm $\times$ 10 cm. As shown in figure 2, each module consists of 4 layers of 58 PRCs each, arranged in orthogonal directions so that two dimensional directional reconstruction is possible. The separation between two successive layers is filled with 15 cm thick concrete. Two layers of PRCs in the same plane are separated by 50 cm allowing the track reconstruction with an accuracy of $6^{o}$ in the projected plane. There are 15 layers of concrete blocks above the top PRC layer to achieve a total thickness of 550 $g/cm^{2}$ giving an energy threshold of 1 GeV for vertical muons. The arrangement of blocks is in form of inverted pyramids to achieve a threshold of 1 $(sec \theta)$ for muons incident at a zenith angle $\theta$.


Angular reconstruction is carried out by checking muon hits below a given layer. As shown in figure 3, a total of 7 angles can be reconstructed in one plane in one direction. This gives a total of 15 bins (2 $\times$ 7 + 1 vertical) in a given plane. Since muon tracking is done in two orthogonal planes, we get a total of 15 $\times$ 15 = 225 directional bins (Figure 4). Since the outer bin contains information on larger angles, which are difficult to model and are heavily influenced by local atmospheric effects, that information is discarded in our analysis. After discarding the outer bins, the total number of bins is 169. These bins are then grouped as shown in Figure 4, so that similar muon counts are obtained in different directions and each bin has enough statistical accuracy for further analysis.

\begin{figure}[h] \begin{center}
\includegraphics[width=3.0in,height=3in]{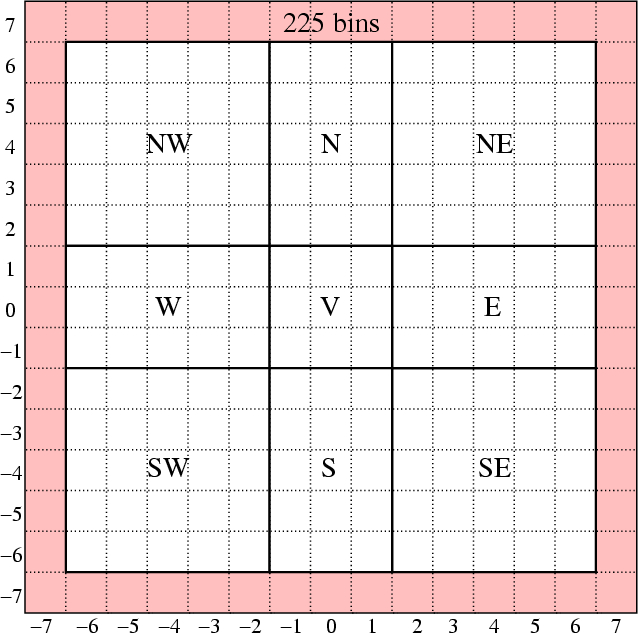}
\caption{\label{fig4} Direction bins in a module}
\end{center} \end{figure} \medskip

\subsection{Computational modeling}

Since the effects we are studying are expected to be small, a large number of muons need to be produced in the simulation to the desired statistical accuracy. Therefore, we simulate 1 billion air showers with proton primaries using CORSIKA 6990 \cite{corsika}, which is a state of the art code to model air showers. CORSIKA parameters were customized based on the location of the experiment. Hadronic interactions in air showers are handled by the low and high-energy interaction models where the transition energy between the two is set to 80 GeV. The low energy hadronic interaction models used for this work were FLUKA 2011, GHEISHA 2002d and UrQMD 1.3, whereas the high energy interaction models were QGSJET-II-03, SIBYLL 2.1 and EPOS 1.99. One combination of models, one low-energy and one high-energy was used to simulated 1 billion air showers. This gives a total of 9 combination of interaction models possible. 

Since the energy threshold of the muon detectors is 1 GeV, muon energies reaching below the threshold were discarded in the simulations. Hadronic thinning option was not used to get good quality data even though simulations became more time intensive. In order to get the desired measurements, we apply a number of quality cuts in our DAQ. Same cuts were applied in simulations. Multiple muons or ``muon bundles" were discarded in order to differentiate muons from hadron penetration due to the punch through effect. Muons incident with higher zenith angles than $60^{o}$ were discarded, as we do in the detector. This also avoids complications with large zenith angle muons related to atmospheric attenuation and other local effects close to the ground level. Showers were randomly incident and muons landing on the detector, crossing the threshold energy were counted in a given direction bin. Since the angular reconstruction depends strongly on the detector geometry, a very detailed survey of the entire setup of PRCs was carried out and incorporated into the simulations. 

\subsection{Data analysis}
PRCs are calibrated once every 24 hours and their pulse width is stored for analysis as the PWA (Pulse Width Analysis) data stream. The amplitude of this PWA curve is directly proportional to the charge deposited in the PRC, which is then converted to muon counts. Each PRC has a different gain and this PWA data is used for calibration before doing physics analysis. The daily mean values in different directions is shown below in Table 1. The flux from vertical direction is highest because of the largest solid angle and least atmospheric attenuation. The East-West effect can also be clearly seen here showing higher muon flux incident from the West than from East.

We also studied the variation of this distribution as a function of time. Figure 5 shows the values of different angular bins as a function of time. Maximum variation observed is 0.3\% over the period of 14 months (Jan 1, 2012 - March 1, 2013). It is safe to conclude that the data is very stable and our analysis does not suffer from any selection bias. Since the total number of muons produced by each model varies, we have normalized the data such that we count the \% of muons falling in each bin, so that experimental data be compared to simulations.

\begin{table}[ht]
\caption{Angular distribution of muons, January 1, 2012. The values here are the 24 hour mean rate in Hz. Please refer to Figure 4 for details on the direction bins.}
\begin{tabular}{c c c }
$NW$ & $N$ & $NE$\\
\hline
$241.058$ &	 $317.958$ & $213.136$\\ 
$337.174$ & $524.096$ & $294.078$\\
$241.311$ & $319.583$ & $215.043$\\
\hline
$SW$ & $S$ & $SE$\\
\end{tabular}\\
\end{table}  

\begin{table}[ht]
\caption{Normalized angular distribution of muons, January 1, 2012. The values represent the \% muons falling in each bin.}
\begin{tabular}{c c c }
$NW$ & $N$ & $NE$\\
\hline
$8.92$ &	 $11.76$ & $7.88$\\ 
$12.47$ & $19.39$ & $10.88$\\
$8.93$ & $11.82$ & $7.95$\\
\hline
$SW$ & $S$ & $SE$\\
\end{tabular}\\
\end{table}  

\begin{figure}[h] \begin{center}
\includegraphics[width=6.0in,height=5in]{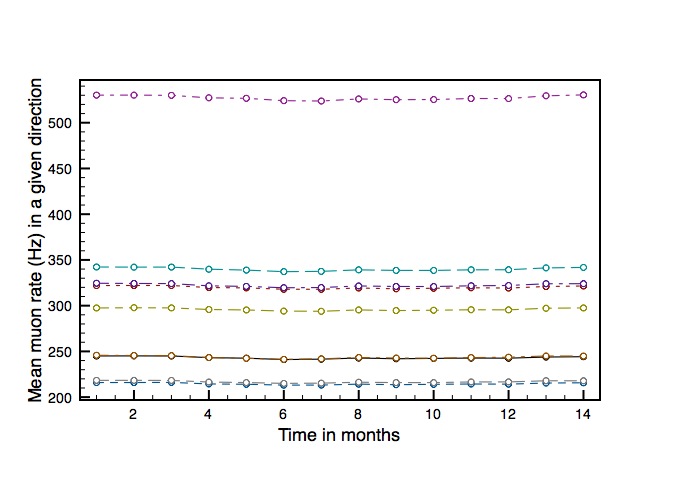}
\caption{\label{fig1} Variation of muon angle rate over a period of 14 months}
\end{center} \end{figure} \medskip

\begin{table}[ht]
\caption{Simulated angular distribution of muons with SIBYLL FLUKA. The values represent the \% muons falling in each bin.}
\begin{tabular}{c c c }
$NW$ & $N$ & $NE$\\
\hline
$8.94$ &	 $11.79$ & $7.85$\\ 
$12.42$ & $19.44$ & $10.92$\\
$8.95$ & $11.85$ & $7.99$\\
\hline
$SW$ & $S$ & $SE$\\
\end{tabular}\\
\end{table}

\begin{table}[ht]
\caption{Simulated angular distribution of muon charge ratio, $ R_{\mu} = (\mu^{+}/\mu^{-}) $, with SIBYLL FLUKA.}
\begin{tabular}{c c c }
$NW$ & $N$ & $NE$\\
\hline
$1.27$ &	$1.22$ & $1.19$\\ 
$1.26$ & $1.21$ &  $1.19$\\
$1.28$ & $1.23$ &  $1.20$\\
\hline
$SW$ & $S$ & $SE$\\
\end{tabular}\\
\end{table}  

\section{Results}
After applying the quality cuts as described in the previous section, muon distributions from different cases were obtained as recorded by the detector. The total number of the detected muons varied depending on the choice of hadronic interaction models, as seen in Figure 6. The lowest number was obtained with the QGSJET-II and FLUKA combination giving $1.96\times10^{7}$ muons and maximum with two combinations, of EPOS and GHEISHA, and QGSJET-II and GHEISHA giving $2.3\times10^{7}$ muons. Figure 5 shows the total values obtained from each model. The maximum difference between the total number of muons detected is 17\%. 

\begin{figure}[h] \begin{center}
\includegraphics[width=6.0in,height=5.0in]{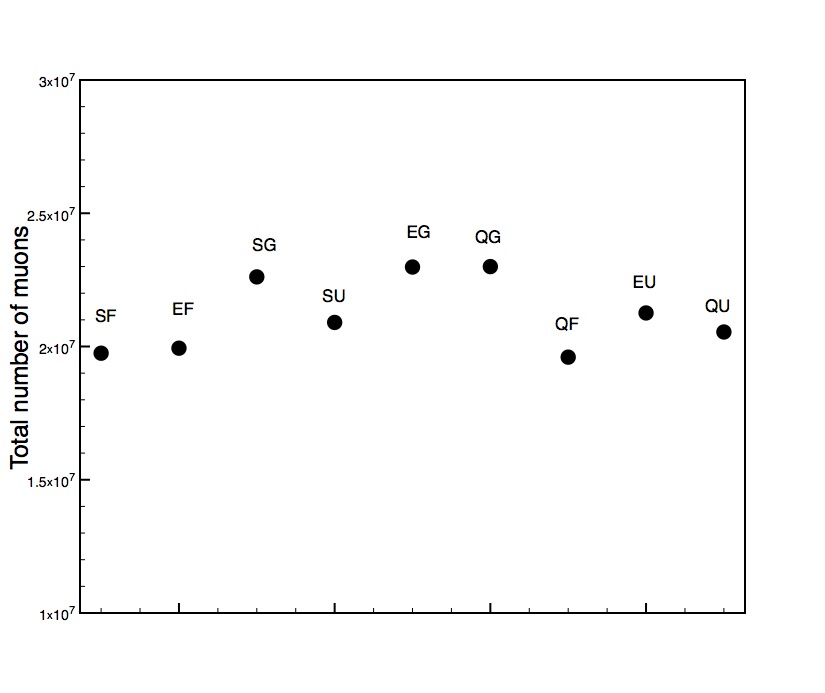}
\caption{\label{fig5} Total number of muons for different combinations of hadronic interaction models. Q is QGSJET-II-03, S is SIBYLL 2.1, E is EPOS 1.99, F is FLUKA, G is GHEISHA and U is UrQMD.}
\end{center} \end{figure} \medskip

The angular distribution of mean primary energies producing muon signal in our detector is also shown in Table 14. The lowest mean energy is incident from the vertical bin because of low threshold and least amount of atmospheric attenuation. East-West asymmetry in primary energies is also clearly visible. Since all the mean energies here are close to the 80 GeV model transition value, both high and low-energy hadronic interaction models play an important role in producing the muon distribution. 

One can clearly see in Tables 1, 2 and 3, the East-West asymmetry in the muon flux in 9 directional bins. Geomagnetic cutoff rigidity plays a major role in shaping this distribution. Atmospheric attenuation is another factor which gives less number of muons for larger zenith angles. Other effects due to the physics of hadronic interactions are not large and hence can not be identified from such plots. 

Data is divided in bins corresponding to the muon detector \cite{hayashi} bins. The detector consists of four orthogonal layers of 58 gas proportional counters each. Details of the detector and binning are described in detail elsewhere \cite{hayashi}\cite{nonaka}. Figure 5 shows the angular distribution binned in 9 direction bins. With $\sim 10^{6}$ muons in each bin, the statistical error is of the order of 0.1\%. This puts us in a good position to compare the results from different combination of models. 

In order to compare different models, all the results were normalized to $2\times10^{7}$. Three low-energy and three high-energy hadronic interaction models gave a total of nine combinations. The \% difference obtained by comparing two set of models is plotted and displayed in the following figures. The maximum difference seen is $\sim 2\%$, which is of high statistical significance ($\sim$ 0.1\%) provided the number of muons in each bin ($\sim$ $10^{6}$). The differences between different combination of models are distinct and clearly displayed in Tables 5 to 13 . 

\begin{table}[ht]
\caption{\% difference between SIBYLL FLUKA and EPOS FLULA.}
\begin{tabular}{c c c }
$NW$ & $N$ & $NE$\\
\hline
$0.51$ &	$0.23$ & $0.55$\\ 
$0.17$ &  $0.03$ & $0.30$\\
$0.64$ &  $0.30$ & $0.58$\\
\hline
$SW$ & $S$ & $SE$\\
\end{tabular}\\
\end{table}  

\begin{table}[ht]
\caption{\% difference between SIBYLL FLUKA and QGSJET FLULA.}
\begin{tabular}{c c c }
$NW$ & $N$ & $NE$\\
\hline
$0.34$ &	$0.25$ & $0.41$\\ 
$0.48$ &  $0.03$ & $0.21$\\
$0.37$ &  $0.01$ & $0.51$\\
\hline
$SW$ & $S$ & $SE$\\
\end{tabular}\\
\end{table}  

\begin{table}[ht]
\caption{\% difference between EPOS FLUKA and QGSJET FLULA.}
\begin{tabular}{c c c }
$NW$ & $N$ & $NE$\\
\hline
$0.86$ &	$0.49$ & $0.98$\\ 
$0.65$ &  $0.06$ & $0.52$\\
$1.03$ &  $0.29$ & $1.10$\\
\hline
$SW$ & $S$ & $SE$\\
\end{tabular}\\
\end{table}  

\begin{table}[ht]
\caption{\% difference between SIBYLL GHEISHA and EPOS GHEISHA.}
\begin{tabular}{c c c }
$NW$ & $N$ & $NE$\\
\hline
$0.44$ &	$0.50$ & $0.56$\\ 
$0.54$ &  $0.37$ & $0.75$\\
$0.62$ &  $0.40$ & $0.58$\\
\hline
$SW$ & $S$ & $SE$\\
\end{tabular}\\
\end{table}  

\begin{table}[ht]
\caption{\% difference between SIBYLL GHEISHA and QGSJET GHEISHA.}
\begin{tabular}{c c c }
$NW$ & $N$ & $NE$\\
\hline
$0.59$ &	$0.55$ & $0.54$\\ 
$0.54$ &  $0.45$ & $0.63$\\
$0.68$ &  $0.38$ & $0.64$\\
\hline
$SW$ & $S$ & $SE$\\
\end{tabular}\\
\end{table}  

\begin{table}[ht]
\caption{\% difference between SIBYLL GHEISHA and QGSJET GHEISHA.}
\begin{tabular}{c c c }
$NW$ & $N$ & $NE$\\
\hline
$0.15$ &	$0.43$ & $0.26$\\ 
$0.10$ &  $0.72$ & $0.27$\\
$0.68$ &  $0.29$ & $0.61$\\
\hline
$SW$ & $S$ & $SE$\\
\end{tabular}\\
\end{table}  

\begin{table}[ht]
\caption{\% difference between SIBYLL UrQMD and SIBYLL UrQMD.}
\begin{tabular}{c c c }
$NW$ & $N$ & $NE$\\
\hline
$0.77$ &	$0.59$ & $0.55$\\ 
$0.49$ &  $0.38$ & $0.51$\\
$0.57$ &  $0.57$ & $0.63$\\
\hline
$SW$ & $S$ & $SE$\\
\end{tabular}\\
\end{table}  

\begin{table}[ht]
\caption{\% difference between SIBYLL UrQMD and QGSJET UrQMD.}
\begin{tabular}{c c c }
$NW$ & $N$ & $NE$\\
\hline
$0.32$ &	$0.48$ & $0.58$\\ 
$0.57$ &  $0.69$ & $0.68$\\
$0.49$ &  $0.41$ & $0.51$\\
\hline
$SW$ & $S$ & $SE$\\
\end{tabular}\\
\end{table}

\begin{table}[ht]
\caption{\% difference between EPOS UrQMD and QGSJET UrQMD.}
\begin{tabular}{c c c }
$NW$ & $N$ & $NE$\\
\hline
$1.12$ &	$1.09$ & $1.15$\\ 
$1.07$ &  $1.08$ & $1.21$\\
$1.06$ &  $0.00$ & $1.16$\\
\hline
$SW$ & $S$ & $SE$\\
\end{tabular}\\
\end{table}


\begin{table}[ht]
\caption{Primary proton mean energy (GeV) of detected muons in 9 direction bins calculated with SIBYLL FLUKA}
\begin{tabular}{c c c }
$NW$ & $N$ & $NE$\\
\hline
$77$ & $73$ & $93$\\ 
$67$ & $64$ & $82$\\
$77$ & $72$ & $91$\\
\hline
$SW$ & $S$ & $SE$\\
\end{tabular}\\
\end{table}

\section{Discussion}

Muons are the messengers of hadronic interactions in the upper atmosphere. We have explored how the angular distribution of muons changes with the use of different hadronic interaction models in air shower simulations. As discussed earlier, there are subtle differences in the physics behind these models. A detailed description of the physical bases of the Monte Carlo codes is available in the literature \cite{tanguy1}\cite{tanguy2}\cite{sibyll}\cite{qgs}. Here, we briefly describe the three high-energy models. 

EPOS 1.99 is compatible with high-energy experiments such as RHIC and SPS, but it produces larger number of muons, when compared to air shower experiments \cite{tanguy1}. It is purely based on quantum mechanics with multiple scattering approach based on patrons and strings. It calculates the cross sections and particle production using energy conservation. Other models, however, do not conserve energy for the calculation of cross sections \cite{tanguy2}. It also accounts for the non-linearities associated with high density effects in heavy ion collisions. The model is used both in air shower experiments as well as in accelerator based experiments. It produces deeper showers compared with other models due to smaller cross sections. It also predicts much lower multiplicity at midrapidity compared with LHC data and other models.

SIBYLL 2.1 is based on the DPM picture in which, a nucleus consists of a quark and diquark, the Lunch Monte Carlo algorithms and the mijijet model \cite{sibyll}. The Glauber scattering theory is used to calculate the total interaction cross sections. For inelastic cross sections and multiplicities, it has a better compatibility with collider and fixed target experiments, compared to its earlier version. However, it overproduces $\pi^{-}$ when compared to fixed target experimental data at FNAL. When it comes to pseudorapidity, it also overestimates particle production at lower energies in the forward region, which is more important for cosmic rays and therefore for this work. It also produces less number of lower energy muons as compared to QGSJET-II-03 and EPOS 1.99. This model is expected to improve by implementing the full Glauber model for nucleus-nucleus interactions

In QGSJET-II-03 (Quark Gluon String Jet model), the hadronic interactions are modeled in the RFT (Reggeon Field Theory) framework, and non-linear effects in nuclear interactions are taken into account in great detail \cite{qgs}. When parton density becomes high, patron cascades strongly overlap and interact with each other, there are shadowing effects and also there is a saturation of parton production. All such effects are carefully taken into account in QGSJET-II-03. The interactions are described as multiple scattering processes, where many parton cascades develop in parallel. Individual scattering contributions are described as parton exchanges. A major theoretical shortcoming of this model is that it does not taken into account the energy-momentum correlations between multiple scattering processes at amplitude level. When compared to TOTEM data, the cross sections rise is higher than observed in the experiment. 

These differences are relatively easier to observe in collider experiments due to controlled nature of the experiment. However, in case of cosmic ray experiments, one can observe a combination of multiple effects described above, which are very difficult to decouple. Although, the differences in final distribution can ultimately reveal the difference in models. 

\section{Summary}
We have modeled the angular distribution of cosmic ray muons using the CORSIKA Monte Carlo simulator. We found that CORSIKA is a very reliable simulator and can accurately reproduce the observed angular distribution of cosmic ray muons. Different combinations of hadronic interaction models were used to explore their effects on the angular distribution. We found that the total number of muons produced by different interaction models differ considerably (17\%). However, upon normalization, the angular distribution does not vary within numerical and systematic uncertainties. This behavior can be ascribed to our understanding of the hadronic interactions at low energies. Table 3 shows that the mean energy of primary proton in different directions is of the order of $\sim$ 100 GeV. This is value is very close to the 80 GeV transition energy between the high and the low energy interaction models. This gives us a unique opportunity to explore both types of models with experimental data. We conclude from this study that the physics of hadronic interactions at these energies is very well understood and have been implemented very efficiently in various event generators as well as in CORSIKA. 

\section{acknowledgement}
The author acknowledges support by the DAE fellowship to carry out this research at the Tata Institute of Fundamental Research in Mumbai and Ooty. We thank S. Gupta, B. Hariharan, Y. Hayashi, P. Mohanty, S. Morris, P. Nayak and the engineering team in Ooty for discussions and technical support.

\end{document}